\documentstyle[psfig]{mn}

\newif\ifAMStwofonts
\ifoldfss
  \ifCUPmtlplainloaded \else
    \NewTextAlphabet{textbfit} {cmbxti10} {}
    \NewTextAlphabet{textbfss} {cmssbx10} {}
    \NewMathAlphabet{mathbfit} {cmbxti10} {} % for math mode
    \NewMathAlphabet{mathbfss} {cmssbx10} {} %  "   "    "
  \fi
  \ifAMStwofonts
    \ifCUPmtlplainloaded \else
      \NewSymbolFont{upmath} {eurm10}
      \NewSymbolFont{AMSa} {msam10}
      \NewMathSymbol{\upi}     {0}{upmath}{19}
      \NewMathSymbol{\umu}     {0}{upmath}{16}
      \NewMathSymbol{\upartial}{0}{upmath}{40}
      \NewMathSymbol{\leqslant}{3}{AMSa}{36}
      \NewMathSymbol{\geqslant}{3}{AMSa}{3E}

    \fi
  \fi
\fi % End of OFSS

\ifnfssone
  \newmathalphabet{\mathit}
  \addtoversion{normal}{\mathit}{cmr}{m}{it}
  \addtoversion{bold}{\mathit}{cmr}{bx}{it}
  \newmathalphabet{\mathbfit} % math mode version of \textbfit{..}
  \addtoversion{normal}{\mathbfit}{cmr}{bx}{it}
  \addtoversion{bold}{\mathbfit}{cmr}{bx}{it}
  \newmathalphabet{\mathbfss} % math mode version of \textbfss{..}
  \addtoversion{normal}{\mathbfss}{cmss}{bx}{n}
  \addtoversion{bold}{\mathbfss}{cmss}{bx}{n}
  \ifAMStwofonts
    \ifCUPmtlplainloaded \else
      \UseAMStwoboldmath
      \makeatletter
      \new@mathgroup\upmath@group
      \define@mathgroup\mv@normal\upmath@group{eur}{m}{n}
      \define@mathgroup\mv@bold\upmath@group{eur}{b}{n}
      \edef\UPM{\hexnumber\upmath@group}
      \new@mathgroup\amsa@group
      \define@mathgroup\mv@normal\amsa@group{msa}{m}{n}
      \define@mathgroup\mv@bold\amsa@group{msa}{m}{n}
      \edef\AMSa{\hexnumber\amsa@group}
      \makeatother
      \mathchardef\upi="0\UPM19
      \mathchardef\umu="0\UPM16
      \mathchardef\upartial="0\UPM40
      \mathchardef\leqslant="3\AMSa36
      \mathchardef\geqslant="3\AMSa3E
    \fi
  \fi
\fi % End of NFSS release 1

\ifnfsstwo
  \DeclareMathAlphabet{\mathbfit}{OT1}{cmr}{bx}{it}
  \SetMathAlphabet\mathbfit{bold}{OT1}{cmr}{bx}{it}
  \DeclareMathAlphabet{\mathbfss}{OT1}{cmss}{bx}{n}
  \SetMathAlphabet\mathbfss{bold}{OT1}{cmss}{bx}{n}
  \ifAMStwofonts
    \ifCUPmtlplainloaded \else
      \DeclareSymbolFont{UPM}{U}{eur}{m}{n}
      \SetSymbolFont{UPM}{bold}{U}{eur}{b}{n}
      \DeclareSymbolFont{AMSa}{U}{msa}{m}{n}
      \DeclareMathSymbol{\upi}{0}{UPM}{"19}
      \DeclareMathSymbol{\umu}{0}{UPM}{"16}
      \DeclareMathSymbol{\upartial}{0}{UPM}{"40}
      \DeclareMathSymbol{\leqslant}{3}{AMSa}{"36}
      \DeclareMathSymbol{\geqslant}{3}{AMSa}{"3E}
    \fi
  \fi
\fi % End of NFSS release 2

\ifCUPmtlplainloaded \else
  \ifAMStwofonts \else % If no AMS fonts
    \def\upi{\pi}
    \def\umu{\mu}
    \def\upartial{\partial}
  \fi
\fi
\newcommand{\gesim}{\,\raisebox{-0.4ex}{$\stackrel{>}{\scriptstyle\sim}$}\,}
\newcommand{\lesim}{\,\raisebox{-0.4ex}{$\stackrel{<}{\scriptstyle\sim}$}\,}
\def\reference{\par \noindent \hangindent=0.75cm}
\def\deg{$^{\circ}$}
\def\kms{ \mbox{km s$^{-1}$}}
\def\hr{$\rm ^h$}
\def\min{$\rm ^m$}
\def\sec{$\rm ^s$}
\def\amin{$'$}
\def\asec{$"$}
\def\sm{$\sim$}
\def\hi{H$\sc i$ }
\def\mfr{$S_{60}/S_{100}$}
\def\mum{$\mu$m}
\def\dv{$\Delta$V}
\def\12co{$^{12}$CO}
\def\exp21{$\times$10$^{21}$}
\title[$^{12}$CO in the Magellanic Bridge]
  {Detection of Carbon Monoxide within the Magellanic Bridge}
\author[E. Muller, L. Staveley-Smith., W.Zealey]
  {E. Muller,$^{1,2}$L. Staveley-Smith,$^2$ W.J. Zealey$^1$\\
  $^1$University of Wollongong, Northfields Ave. 
  Wollongong, NSW 2500, Australia\\
  $^2$Australia Telescope National Facility, CSIRO, PO Box 76, Epping, N.S.W 1710 Australia}

\date{}

\pagerange{\pageref{firstpage}--\pageref{lastpage}}
\pubyear{2002}

\begin{document}

\maketitle

\label{firstpage}

\begin{abstract}
  The Mopra 22m and SEST 15m telescopes have been used to detect
  and partially map a region of \12co(1-0) line emission
  within the Magellanic Bridge, a region lying between the Large and
  Small Magellanic Clouds.  The emission appears to be embedded
  in a cloud of neutral hydrogen, and is in the vicinity of an IRAS
  source.  The CO emission region is found to have a 60\mum/100\mum\
  flux density ratio typical for \12co(1-0) detections within the SMC,
  although it has a significantly lower \12co brightness and velocity
  width. These suggest that the observed region is of a low
  metallicity, supporting earlier findings that the Magellanic Bridge
  is not as evolved as the SMC and Magellanic Stream, which are
  themselves of a lower metallicity than the Galaxy. Our observations,
  along with empirical models based on SMC observations, indicate that
  the radius of the detected CO region has an upper limit of \sm16 pc.
  This detection is, to our knowledge, the first detection of CO
  emission from the Magellanic Bridge and is the only direct evidence
  of star formation through molecular cloud collapse in this region.
\end{abstract}


\section{Introduction}
The Magellanic Bridge forms a link of primarily neutral hydrogen
between the Small and Large Magellanic Clouds (SMC and LMC). Recent
high-resolution observations of the \hi fraction of the Bridge, using the
Australia Telescope Compact Array (ATCA) and the Parkes Telescope by
Muller et al. (MNRAS, in press), show an intricate morphology, comprising
numerous clumps and filaments across all observed spatial scales.  The
mechanism responsible for the formation of the Bridge is widely
considered to be the gravitational influence of the LMC, and numerical
simulations have shown that its formation was possibly triggered
during a close pass of the Clouds to each other around 200Myr ago (eg
Gardiner \& Noguchi 1996).

The stellar population of the Bridge has been shown to be relatively
young, with an age range of \sm10 to 25 Myr (Grondin, Demers \&
Kunkel, 1992 and Demers \& Battinelli, 1998). The young ages imply
that star-formation within the Bridge is an active process, yet no
previous evidence of star-forming regions through CO emission has been
found.

The Magellanic Clouds, as well as the Magellanic Bridge, have been the
focus for a number of searches of molecular transition lines. From
these searches, a number of different carbon isotopes, as well as
other molecular species have been identified.  It appears that the
Clouds are rather metal deficient, the SMC particularly so (Israel et
al. 1993, and papers of that series). The low CO luminosity of the
SMC, being weak and cold in comparison to Galactic Molecular Clouds,
has been ascribed to a smaller molecular cloud size which may be, in part,
due to a higher UV radiation field causing more thorough
photo-disassociation (Rubio et al., 1991, Israel et al., 1993).  These
authors suggest that the higher UV field of the SMC is the product of
a more active star formation per unit mass than the galaxy, and lower
absorptive dust fraction.  Rubio et al. (1993) note that the
luminosity and metallicity of CO regions within the SMC are similarly
less than those of the Galaxy, although they do not speculate on a
reliable and direct correlation between these two parameters.
Spectral studies of young, early-type stars throughout the SMC
(Rolleston et al., 1993) have shown that the heavy-metal abundance
within this SMC is not homogeneous, and is lower than Galactic values.

A few stars within the Bridge and SMC wing have been found to be more metal
deficient than star-forming regions in the SMC by \sm0.5 dex (Rolleston et al.,
1999).  More recent spectral studies towards the centre of the Bridge
by Lehner et al. (2001) confirm the mismatch of abundances between the
SMC and the Magellanic Bridge, and these authors suggest that the
Bridge comprises a mix of gas from SMC gas, and from its relatively
un-enriched halo.
 
Israel et al. (1993) made extensive searches for CO emission regions
within the SMC towards locations selected from IRAS maps and from
known H$\sc ii$ emission regions.  They found that strong sources of
CO emission were generally associated with regions where the ratio of
60\mum\ and 100\mum\ flux densities (\mfr) were \mfr\lesim1.0, and that
sites of CO emission where \mfr\gesim1.0 were relatively weak. In
general, SMC CO line emissions were found to be weaker and narrower
than those of Galactic CO emission.

At this stage, CO emission of any kind has not previously been reported in the
Magellanic Bridge further east than the molecular cloud N88, which
occupies the north eastern corner of the SMC.  N88 has also been
observed in a number of other molecular lines (Testor et al., 1999).
Smoker et al. (2000) have conducted a search for \12co(1-0) near
the centre of the Magellanic Bridge, towards a region of cold atomic
hydrogen (Kolbunicky \& Dickey 1999).  This target was considered a
likely candidate based on studies by Garwood \& Dickey (1989), who had
found that CO emission regions were occasionally associated with cold
atomic gas.  These SEST observations by Smoker et al.  however, showed
no \12co(1-0) emission down to an RMS of 60 mK.

Maps of CO emission from the tidally affected M81 galaxy have shown
that some CO regions can be associated with tidally extruded \hi
(Taylor, Walter \& Yun, 2001), although the correlation is not
particularly outstanding and the CO appears to only loosely trace the
\hi mass.

CO emission in Tidal Dwarf Galaxies (TDGs) has been found to generally
correlate with regions of high \hi column density by Braine et al.
(2001), although only single pointings were made during this study.
TDGs are thought to condense from the remnant material exported from a
host galaxy during a tidal stripping event (eg Braine et al. 2001).
They may represent a class of objects where star-formation has
preceded significant tidal perturbation. This is the reverse of the
sequence of processes thought to be active in the Magellanic Bridge
and is a relevant benchmark to bear in mind.

This paper presents results from a CO survey of a region in the
Magellanic Bridge using selection criteria based on findings by Israel
et al. (1993), Taylor, Walter \& Yun (2001) and Braine et al. (2000,
2001).

We discuss the selection procedure of candidate CO emission sites in
Section \ref{sec:selection}. Observation techniques are outlined in
Section \ref{sec:obs}. In Section \ref{sec:analysis}, we present the
results and comparisons with \hi data. We discuss the results in
Section \ref{sec:discussion}.
\begin{center}
\begin{table*}
\begin{tabular}{cllccccc}\hline\hline
  Pointing &\multicolumn{2}{c}{Offset}& \hi Col. dens.  &\mfr
  &Obs.time&RMS (after reduction)&S/N\\ 
  &RA&Dec.&cm$^{-2}$&&(min)&(mK)&\\\hline 1 &0.0 & 0.0 &2.7\exp21&0.18
  &157&21&6.5\\ 2 &0.0 & $+$45\asec &2.7\exp21&0.24&48 &34&6.1\\ 3
  &$+$45\asec & $+$45\asec &2.6\exp21&0.23&48 &39&-\\ 4 &0.0 & +90\asec
  &2.5\exp21&0.26&48 &39&3.1\\ \hline \hline
\end{tabular}
\caption{Positions, \hi column densities and \mfr\ values for the
  detected CO cloud, and adjacent positions (See
  Figure~\ref{fig:intintmap}). The reference position is at RA (J2000) =
  01\hr56\min47\sec, Dec. (J2000) =$-$74\deg17\amin41\asec}
\label{tab:pointstats}
\end{table*}
\end{center}

\section{Source Selection}\label{sec:selection}
CO requires a surface on which to form, and in instances where UV flux
may be high enough to cause disassociation, a buffer of absorbing
material is necessary to surround the CO region. As such, candidate CO
regions were selected for this survey primarily through comparisons of
maps of far-IR emission and of \hi column density. In general, 60\mum\
emission across the Bridge is negligible except for a few sparsely
distributed bright peaks, with a maximum brightness of \sm1.5 MJy
sr$^{-1}$. As mentioned previously, the \hi distribution is rather
clumpy with a filamentary component throughout much of the Bridge. The
\hi spectral information used throughout this study has been extracted
from the \hi datacube, presented and discussed in a paper by Muller et
al (MNRAS, in press).  This dataset comprises observations with the Australia
Telescope Compact Array (ATCA) and the Parkes telescope.  The
sensitivity of these observations is \sm15.2 mJy/Beam and a beamwidth
of \sm98\asec.

Candidate regions were those where the 60\mum\ to 100\mum\ flux density
ratio \mfr$<$0.2 were co-incident with a local \hi integrated
intensity maximum. Throughout the Bridge, six candidate regions
satisfying this criteria have been identified. At this time however,
only one of these regions has been investigated for \12co(1-0)
emission.  The other five, un-observed CO candidate sites have lower
\mfr\ ratios, lower 60\mum\ emission, and with one exception, all have
OB associations and IRAS sources within a few tens of parsecs
(projected).  Study of the remaining sites will be the subject of a
future paper.  

\begin{figure*}
  \centerline{ \psfig{file=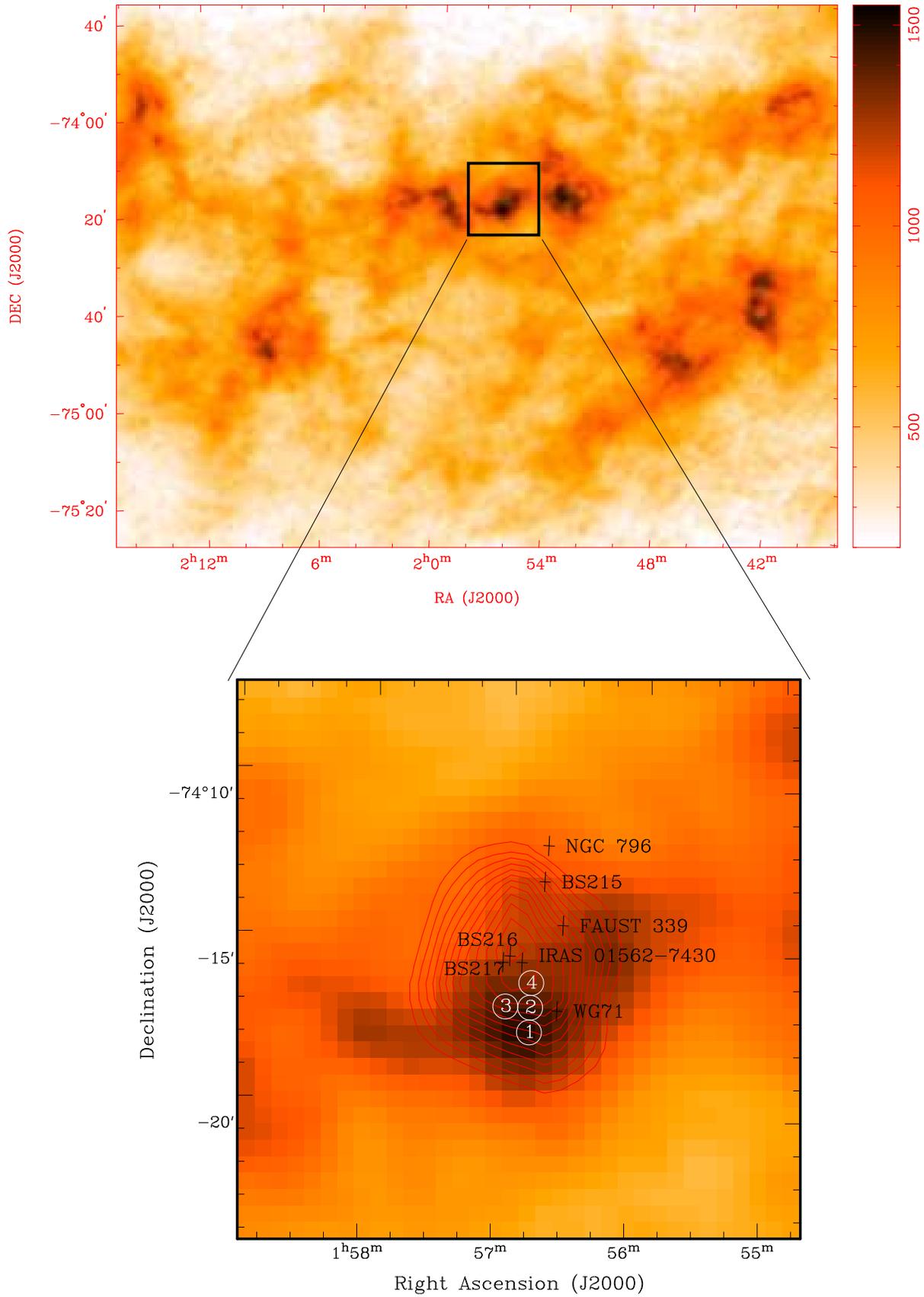,width=16cm}}
\caption{Integrated \hi intensity map the
  Magellanic Bridge, and a blowup of the candidate CO emission area.
  Pointings are shown as circles labelled as pointings 1-4.  Contours
  correspond to the 60\mum\ brightness of 0.1 to 1.20 MJy sr$^{-1}$.
  Crosses define the positions of OB associations, FAUST objects and
  IRAS sources nearby (from catalogues by
  Bica et al., 1995 and Westerlund \& Glaspey, 1971)}
\label{fig:intintmap}
\end{figure*}

\section{Observations}\label{sec:obs}
\subsection{Mopra}
The 22m Mopra telescope \footnote{The Mopra telescope is part of the
  Australia Telescope which is funded by the Commonwealth of Australia
  for operation as a National Facility managed by CSIRO.} was used to
examine one pointing only. These observations were made during the
evening and night of 11$^{th}$ of October 2000, and during the night
of 13$^{th}$ December 2000.

The two SIS receivers of Mopra were tuned to 115.19GHz.  The
correlator was configured to give 64 MHz bandpasses with 1024 channels
for both receivers, resulting in a velocity range of \sm167\kms\ and a
velocity channel spacing of \sm0.163 \kms.

During observing runs, the telescope was calibrated every \sm45
minutes with observations of an ambient temperature source. Reference
observations were made of an area located two minutes south of the
source. Total on-source time was \sm682 minutes.

Observations of N88 were made for calibration purposes. The
\12co(1-0) signal strength of \sm200mK for N88 is 40\% of the brightness
measured by Rubio et al. (1996), and is in general agreement with the
estimated efficiency of the Mopra Telescope (Kesteven M., Priv.comm.).

The pointing corrections for the telescope were made before each \sm6
hour observing session using SiO maser sources.  Corrections were
repeated until accuracy was better than 5''.

All data was reduced following standard procedures, using the ATNF SPC
reduction package.

\subsection{SEST}
Confirmation and mapping observations were conducted with the SEST
Swedish ESO Sub-millimetre Telescope, during the evenings and nights
of the 2001 December 15$^{th}$-17$^{th}$.

The SESIS 100 receiver was tuned to 115.19 GHz, and the correlator was
configured to the High Resolution Spectrometer settings, giving a
45$''$ beamwidth, and a 83.6MHz bandpass of 2000 channels.  This
yields a velocity range of \sm218\kms\ and a velocity channel spacing
of \sm0.109\kms.

The Double Beam Switching (DSW) mode was used for these observations.
This employs a focal plane chopper wheel that rotates at 6Hz to
observe the off-source position, located +2\min27\sec in azimuth away
from the source.

During observing runs, pointing and focus corrections were made every
two hours, using the SiO maser R Doradus (\sm8Jy) where possible,
otherwise R Aquarius was used (\sm5Jy).  Calibration checks were made
every \sm5 minutes.

Four pointings were made at beam-width sized offsets from the primary
position.  Positions, \hi column densities and \mfr\ values for these
pointings are shown in Table~\ref{tab:pointstats}.  The pointings are
numbered in the order in which they were observed.

Figure~\ref{fig:intintmap} shows the position of the candidate
emission region on an integrated intensity map (Muller, et al., MNRAS,
in press).
The sub-image shows the observed positions, nearby OB associations
extracted from an OB catalogue (Bica et al. 1995) and contours of 60\mum\
brightness.

Pointing 1 was observed for a total of 157.1 minutes, on-source time.
The other three positions were observed for 48 minutes each, on-source
time.  System temperatures ranged between \sm500-590 K.

Data were reduced using the XS data reduction package. The raw spectra
had velocity resolution of \sm0.11\kms. This was increased to a final
value of \sm0.33\kms\ during reduction to increase the
\mbox{signal-to-noise}.  Data were scaled to compensate for the
main-beam efficiency using: T$_A$=T$_{MB}$/0.7. All data were baseline
subtracted with a single linear baseline, although in most cases this
was hardly necessary, and a higher order fit did not improve S/N significantly. The resulting RMS values, (after rebinning
across three velocity channels) and the on-source times are shown in
Table~\ref{tab:pointstats}.

\section{Spectra and analysis}\label{sec:analysis}
The primary pointing, and two of three offset pointings at the studied
location yielded positive detections (SEST).

\begin{figure}
  \centerline{\psfig{file=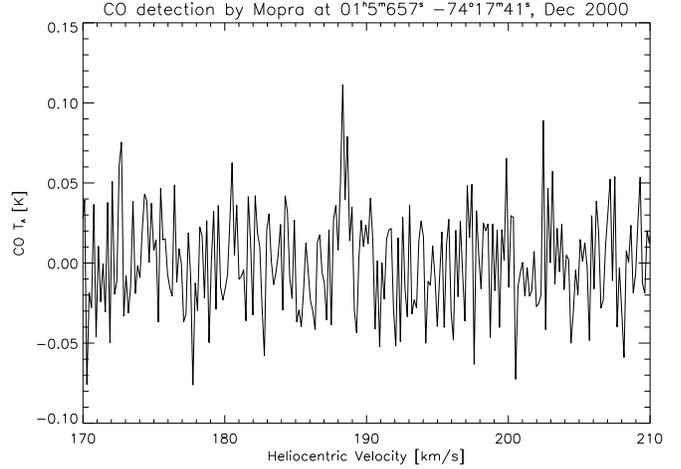,width=9cm}}
\caption{Initial detection made with the ATNF Mopra Telescope.}
\label{fig:mopradect}
\end{figure}

\begin{figure*}
  \centerline{ \psfig{file=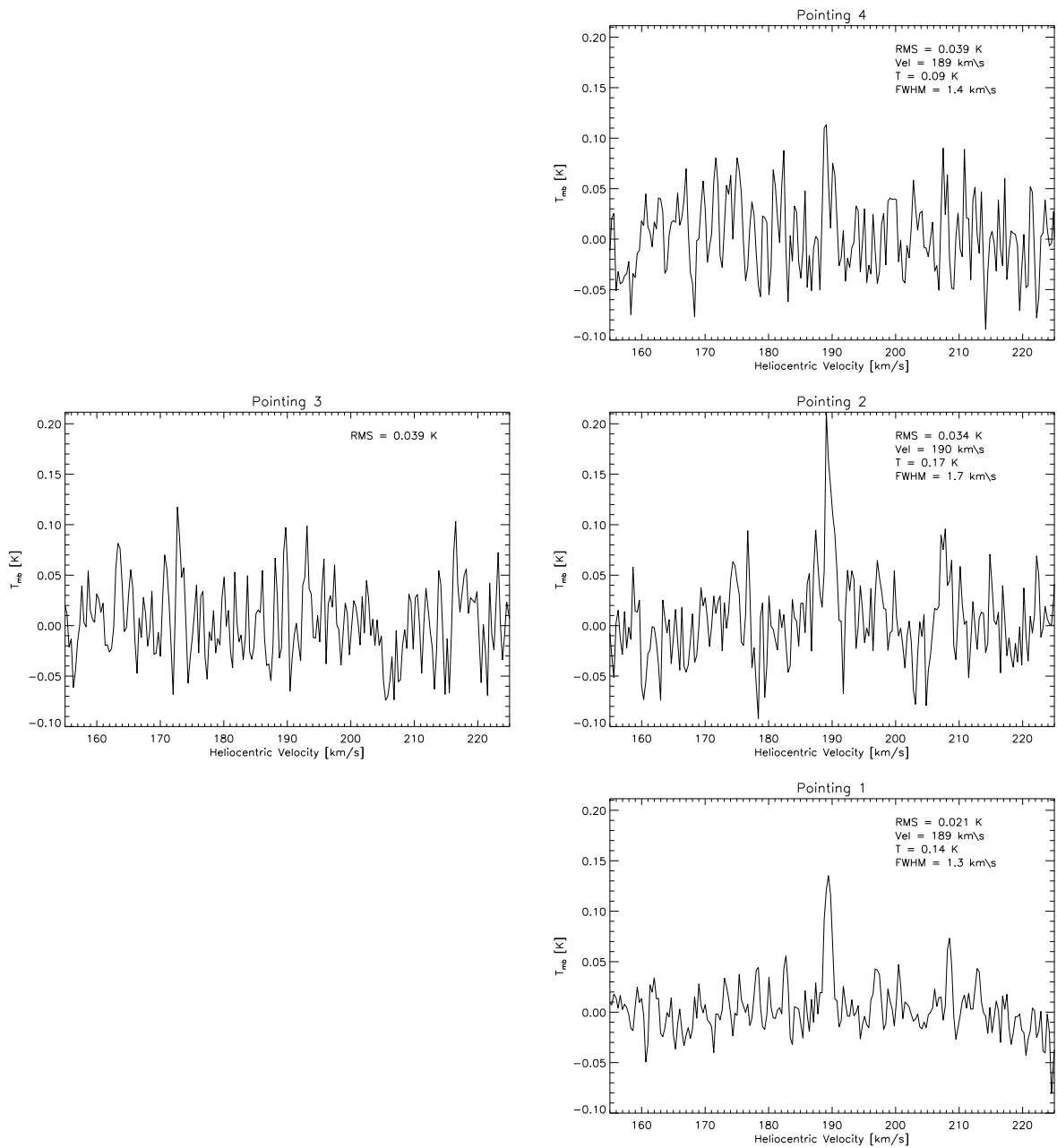,width=16cm,angle=-90}}
\caption{Smoothed spectra over velocities \sm160\kms\ to 220\kms\
  (Heliocentric). Gaussian fit parameters are shown in the top right
  of each plot. See Table~\ref{tab:gaussfits} also.}
\label{fig:spectrazoom}
\end{figure*}

The initial detection made with the Mopra telescope is shown in
Figure~\ref{fig:mopradect}, and has a S/N of \sm3.5,  while spectra
obtained with the SEST are shown in Figure~\ref{fig:spectrazoom} for
each of the pointings 1-4.
Table~\ref{tab:gaussfits} summarises the parameters for gaussian fits
made to spectra from pointings 1,2 \& 4, as well as integrated intensity
determined from the fit parameters.  The analysis package 'XS' was to
make the gaussian fitting.

The detection at Pointing 4 has a S/N of \sm3.1, and is not a
significant detection on its own, however positive detections made
nearby improves its legitimacy.  Pointing 3 did not yield a convincing
detection of the \12co(1-0) line.

\begin{center}
\begin{table}
\begin{tabular}{ccccc}\hline\hline
  Pointing&V$_o$&I$_{CO}$\\ &\kms&mK &\kms
  &K$^.$\kms\\ \hline 
1&189.4&140&1.3&0.19\\ 
  2&189.5&170&1.7&0.31\\
  4&189.3&90&1.4&0.13\\ \hline\hline
\end{tabular}
\caption{Parameters of Gaussian fit to smoothed data for pointings 1,2 and
  4.  Central Velocity is consistent to within $\pm$0.2\kms, although
  line width (\dv, FWHM) varies by up to \sm23\%}
\label{tab:gaussfits}
\end{table}
\end{center}

We can see from Fig.~\ref{fig:intintmap} that pointing 1 corresponds to
the highest local \hi column density.  This point appears to be on the
rim of an annular, or crescent-shaped \hi filament and has a
\mfr$=$0.18, although this parameter peaks slightly north of the \hi
integrated intensity maximum at the IRAS catalogue object, IRAS
0152-7430 (\mbox{01\hr56\min57\sec -74\deg14\amin14\asec, J2000}). Two
OB populations are located at almost the same position as the IRAS
object, comprising either one association with a more centrally
condensed population, or two superimposed OB associations.  These
associations are labelled BS216 and BS217 (Bica et al. 1995). A third
OB association is found to the west of pointing 2. This association is
labelled WG71 (\mbox{1\hr56\min35\sec -74\deg17\amin1\asec,}
 Westerlund \& Galespey, 1971).

\begin{figure}
  \centerline{\psfig{file=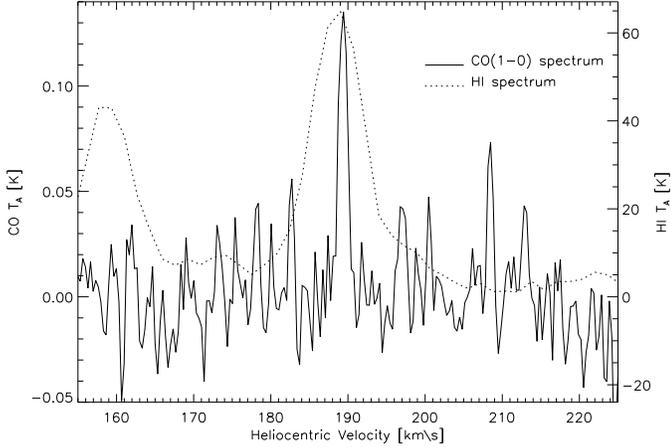,width=9cm}}
\caption{Smoothed CO spectrum at pointing 1, overlaid on \hi spectra
  at same location.  {\em Dotted line and right axis:} \hi spectrum.
  {\em Solid line and left axis:} CO spectrum.  This figure shows the
  excellent match in velocity between the peaks of the spectra,
  suggesting that the CO cloud is embedded within an \hi cloud.}
\label{fig:CO+HI}
\end{figure}

The detected CO spectrum from pointing 1 is shown in
Fig.~\ref{fig:CO+HI}, with the \hi spectrum at the same location
overlaid (Muller et al., MNRAS, in press).  The peak velocity of the CO
emission coincides with that of the \hi emission, indicating that the
CO emission region is enclosed by an \hi cloud which may be offering
some protection from ambient UV flux.  The width of the \hi line at
this location is \sm10 \kms.  More mapping will be required to
ascertain the extent of the CO mass and the degree of correlation of
the \hi and CO emission regions.

\begin{figure}
  \centerline{\psfig{file=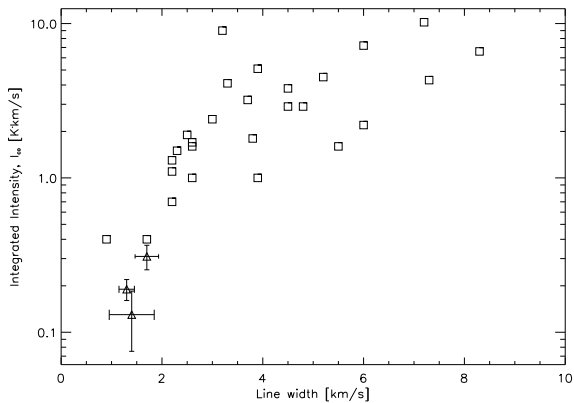,width=8cm}}
\caption{Integrated CO intensity against line-width after Israel et
  al. (1993) (Figure 6b). \12co detections within the SMC are shown as
  squares, with data from this work are shown as triangles with error
  bars.  These new detections are consistent with the previous trend
  between CO line width and CO intensity.}
\label{fig:israel}
\end{figure}

We have plotted the integrated CO intensity, L$_{CO}$ as a function of
line-width in Figure~\ref{fig:israel}, along with those for detections
made in the SMC by Israel et al. (1993). This plot highlights the
relative weakness and the low velocity dispersion of the CO detection
in the Magellanic Bridge. Visually, the plot also suggests that the
Bridge detections fit quite well into a correlation between velocity
width and I$_{CO}$ for SMC detections. The value of I$_{CO}$ plotted
by Israel et al. is a direct sum, whereas our value of I$_{CO}$ is
determined from a gaussian fit. However, since the emission line
appears to have no velocity structure, the difference between the
methods is considered to be negligible.  The errors for this plot were
determined using the algorithms developed by Landman, Roussel-Dupr\'e
\& Tanigawa, (1982), and Barranco \& Goodman (1998).

\section{Discussion}\label{sec:discussion}
\mbox{From the observations around 01\hr56\min48\sec $-$74\deg16\amin56\asec
(J2000),} three separate locations have shown a
positive $^{12}$CO(1-0) detection.  The pointing from which the CO
emission is absent does not appear to be significantly different to
the other three positions in terms of \mfr\ and \hi integrated
intensity.

\begin{figure}
  \psfig{file=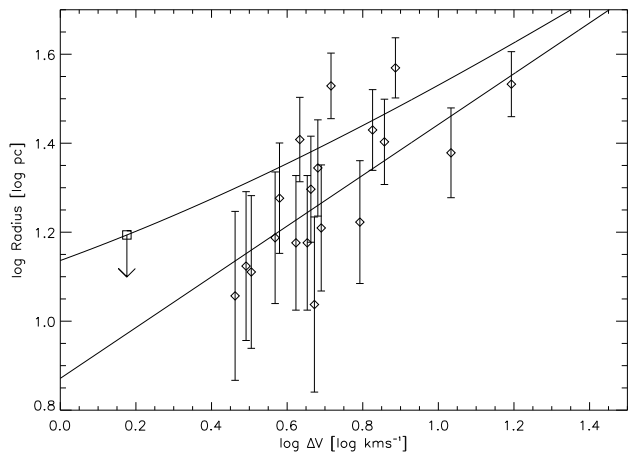,width=9cm}

\caption{Plot of the dependence of CO emission region radius (pc) on velocity width (\dv) for SMC CO emission regions {\em (diamonds)}.
  The straight {\em lower line} is a fit to the SMC dataset collated
  by Rubio et al.  (1993). This figure differs to that of figure 1 of
  that paper, in that the axes are reversed and the radius is treated
  as the dependent variable. The upper {\em curved line}
  represents an upper limit to the log-log fit, and is half the size
  of the beam used in the SMC observations. The square marks the upper
  radius limit of the Magellanic Bridge detection at pointing 2
  (\dv=1.7 \kms).  Error bars are one half the beamwidth for the
  observations ($\approx$6.3pc)}
\label{fig:rubiodat}
\end{figure}

Rubio et al. (1993) were able to formulate empirical relations from
their studies of SMC CO emission regions. These relations allow the
velocity dispersion, luminosity, virial mass and H$_2$/CO factor to be
estimated from a measurement of the radius of the CO emission region.
We estimate a new fit to only the SMC data from the paper by Rubio et
al. (1993), and find a relation of: \mbox{$\log(R)=0.57\log$\dv$+0.87$.} The
dataset from Rubio et al. (1993) is shown in Fig.~\ref{fig:rubiodat}
along with the fit determined here.  We have included an upper limit
to the log-log fit, based on the beamsize used for those observations,
from which we estimate an upper limit to the radius of the Magellanic
Bridge CO emission region of \sm16 pc.  We see from
Fig.~\ref{fig:spectrazoom} that the emission region extends across
multiple beam-widths, where one beamwidth subtends \sm10pc.  From the
relative amplitudes of each of the positive detections, We suggest
that the centre of the CO emission cloud is southwest of pointing 2,
and is perhaps co-incident with the association WG71. We should bear
in mind that the Magellanic Bridge has a slightly lower metal
abundance than the SMC by \sm 0.5 dex (Rolleston et al.  1993), and we
would expect this to reduce the size of the emission region (Rubio et
al., 1993).

CO has not been detected in emission for any galaxy having a
12+log(O/H)\lesim7.9 (Taylor, Kobulnicky \& Skillman, 1998). Taking a
mean value for 12+log(O/H) for the SMC of \sm7.98 (Pagel et al, 1987), along
with the knowledge that isolated measurements of the Magellanic Bridge show
an under-abundance relative to the SMC by 0.5 dex (Rolleston et al.
1999), we see that this may be the first successful detection of CO in
emission where 12+log(O/H)\lesim7.9. We should note that the
estimates of 12+log(O/H) by Pagel et al. were made from studies of
H$\sc ii$ regions, and range between 7.84 and 8.16. More importantly,
the estimates of the 12+log(O/H) for the Magellanic Bridge were made
at a location a few degrees east from the location of the CO
detection, and Rolleston et al. (1993) have shown that the metallicity
of the SMC is not homogeneous. It is probable then, that the metallicity
of the Magellanic Bridge is equally inhomogeneous.

A mean conversion factor for
X=$N(H_2)/I_{CO}$=$(120\pm20)\times10^{20}$cm$^{-2}$ was determined by
Israel (1997) from analysis of various CO emission regions in the SMC.
It is interesting to apply this factor to the CO emission detected in
the Magellanic Bridge.  Table~\ref{tab:xfact} shows the predictions of
the H$_2$ column densities using this transformation. We also see from
this table that the ratio of M(H$_2$)/M(\hi) is significantly larger
than the mean M(H$_2$)/M(\hi) for the SMC. To maintain a consistent
M(H$_2$)/M(\hi) between the Bridge and SMC, a lower X value is
required for the Bridge. This is as we would expect for a
lower-metallicity environment.

\begin{center}
\begin{table}
\begin{tabular}{ccc}\hline\hline
Pointing&N(H$_2$)&M(H$_2$)/M(\hi)\\
        &$\times10^{21}$cm$^{-2}$&\\\hline
1&2.3&1.7\\
2&3.7&2.8\\
4&1.6&1.3\\
\hline
SMC& - &0.2 (Mean)\\
\hline
\end{tabular}
\caption{Estimates of H$_2$ column density using the SMC factor
  $X=N(H_2)/I_{CO}=(120\pm20)\times10^{20}cm^{-2}$. The predicted H$_2$ column
  density is comparable to that of the \hi column density. This is an
  interesting result in the context of the SMC, where the
  M(H$_2$)/M(\hi) ratio is significantly smaller (Israel, 1997).}
\label{tab:xfact}
\end{table}
\end{center}

The general nature of \hi within the Bridge is clearly turbulent (see
Muller, et al., MNRAS, in press) and the \hi in the neighbourhood of these
detections has a clear bimodal profile, with the two peaks separated by a
lower density rift of \sm 30\kms\ (Heliocentric).  The CO detections
made here correspond to the velocity of the receding peak. The velocity
bifurcation continues into the SMC, and about one third of the way to
the LMC (Muller et al., MNRAS, in press).  Rubio et al. (1993) noted that all the
CO detections in the SMC were located on the approaching velocity
peak, although any reason or significance for a preference is unclear.
A more detailed study of this area shows that it cannot be interpreted
as an expanding \hi shell (Muller, et al. MNRAS, in press).

Studies of the CO emission from Tidal Dwarf Galaxies and tidal
remnants by Braine et al. (2000) and Braine et al. (2001) have shown
that molecular gas is observable within the remnants of tidal
interaction and these authors suggest that the CO regions form after
the tidal stripping event. Typical line-widths and I$_{CO}$ values
from these observations of the \12co(1-0) transition are \sm15 - 66
\kms\ and 0.1 - 2 K\kms. However, that sample has an oxygen abundance
range of 12+log(0/H)\sm8.2 - 8.6, which is in the upper range of that
found from a study of dwarf galaxies including the SMC and the LMC
(Israel, Tacconi \& Baas, 1995). The high metallicities of these type of
galaxies are considered to be the remnants of their original host galaxies. As
the SMC is thought to be the original host for the matter drawn out
into the Magellanic Bridge, we would expect the line-widths of the Bridge CO
emission regions to be similar to those of the SMC and Figure~\ref{fig:israel} shows that this is indeed the case.  Braine et
al. (2000) selected CO candidate emission regions on the basis of high
\hi column density. They also concluded that the spatial co-incidence
of \hi clumps and CO regions suggested that the CO region had formed
after the tidal extraction event.  From the lack of information of the
spatial distribution of the CO emission region in the Bridge, it is
difficult to say whether the CO emission region coincides exactly
with the nearby \hi clump. At this stage however, the spatial and
velocity positions of \hi clumps and the CO emission region appear to
correspond. A future paper will examine this aspect in more detail.

At this stage then, it appears then that the narrow CO line-widths as
measured in the Magellanic Bridge are not representative of those of
observed TDGs, although we should bear in mind the differences in the
progenitor galaxy types, and we should be careful in drawing parallels
with this context.

\section{Conclusions}\label{sec:conclusions}
We have detected a $^{12}$CO(1-0) emission region within the
Magellanic Bridge.  The line width and integrated intensity is
compatible with other CO detections within the SMC, although it is
somewhat weaker and cooler, implying a lower metallicity.  The
findings at this point appear to be consistent with studies of CO
emission regions within the SMC where CO has been found to associated
with high \hi integrated intensity and with \mfr$<$0.2. Estimates made
from log-log model fits to SMC CO observations indicate that the
Magellanic Bridge CO emission region is spatially small, and has a
radius of $<$16 pc.  The CO emission
region has a narrow line-width compared with observations of the SMC
and with other TDGs, lending
support to findings that suggest the Bridge has a lower metallicity
than the SMC. At this stage, a reliable estimation of H$_2$ mass is
not really possible from the small amount of spatial information,
although these issues will be addressed in a future paper.  This
finding confirms for the first time that star formation through
molecular cloud collapse is an active and current process within the
Bridge.

\section*{Acknowledgements}
The Authors would like to thank the referee, Dr. Frank Israel, for his
helpful comments and suggestions for improvements on this paper.
Thanks goes also to Dr. Tony Wong for essential advice on early drafts of
the paper. We would also like to thank the Staff at the SEST facility
and the European Southern Observatory. Finally, we would like to
acknowledge the ATNF and CSIRO for the use of the Mopra Observatory.

\section*{References}
\reference Barranco, J.A., Goodman, A.A.,1998, ApJ, 504, 207B
\reference Braine, J., Duc, P.-A., Lisenfeld, U., Charmandaris, O., Vallego, O., Leon, S., Brinks, E. 2001, A\&A, 378, 51
\reference Braine, J., Lisenfeld, U., Duc, P-A., Leon, S. 2001, Nat., 403, 867
\reference Bica E.L.D., Schmitt H.R. 1995, ApJS, 101, 41 
\reference Demers, S., Battinelli, P., 1998, AJ 115, 154
\reference Gardiner, L., Noguchi, M., 1996, MNRAS, 278, 191
\reference Garwood, R.W., Dickey, J.M., 1989, ApJ, 338, 841
\reference Grondin, L, Demers, S, Kunkel, W.E., 1992, AJ, 103, 4, 1234
\reference Israel, F.P., Johansson, L.E.B., Lequeus, J., Booth, R.S., 
Nyman,
  L.,-A, Crane, et al. 1993, A\&A, 276 , 25
\reference Israel, F.P., Tacconi, L.J., Baas, F., 1995, A\&A, 295, 599
\reference Israel, F.P., 1997, A\&A, 328, 471.
\reference Kobulnickey, H.A., Dickey, J.M., 1999, A\&A, 117, 908
\reference Landman, D.A., Roussel-Dupr'e, R., Tanigawa, G., 1982, ApJ, 261, 732
\reference Lehner, N., Sembach, K.R., Dufton, P.L., Rolleston, W.R.J.,
  Keenan, F.P., 2001, ApJ, 551, 781

\reference Pagel, B.E.J., Edmunds, M.G., Fosbury, R.A.E., Webster,
B.L., 1978, MNRAS, 184, 569
\reference Rolleston, W. R. J., Dufton, P. L., Fitzsimmons, A., Howarth, I. D., Irwin, M. J., 1993, A\&A. 277,10
\reference Rolleston, W.R.J., Dufton, P.L., McErlean, N.D., Venn, K.A., 1999, A\&A, 348
\reference Rubio, M., Garay, G., Montani, J., Thaddeus, P., 1991, ApJ, 368, 173.
\reference Rubio, M., Leuqeus, J., Boulanger, F., 1993, A\&A, 271, 9
\reference Rubio, M., Lequeus, J., Boulanger, F., Booth, R.S., Garay,
G., de Graauq, Th., Israel, F.P., et al. 1996, A\&A supp. ser. 118, 263
\reference Smoker, J.V., Keenan, F.P., Polatidis, A.G., Mooney, C.J.,
  Lehner, N., Rolleston, W.R.J., 2000, A\&A, 363, 451
\reference Taylor, C.L., Walter, F.,Yun, M.S., 2001, ApJ, 562, L43
\reference Taylor, C.L., Kobulnickey, H.A., Skillman, E.D., 1998, AJ,
116, 2748
\reference Testor, G., Rola, C.S., Whiing, A.B., 1999, In: Y.-H.Chu Suntzeff,
N.B., Hesser, J.E., Bohlender, D.A. ed., IAU \#190, New Views of
  the Magellanic Clouds. 
\reference Westerlund, B.E., Glaspey, J., 1971, A\&A, 10, 1

\bsp
\label{lastpage}

\end{document}